\documentclass[fleqn,twoside]{article}
\usepackage{gc,amsfonts,bezier,epsf}

\def\GR{general relativity}
\def\sph{spherically symmetric}
\def\ssph{static, spherically symmetric}

\def\bh{black hole}

\def\qua{\fract 14}
\def\mn{_{\mu\nu}}
\def\MN{^{\mu\nu}}
\def\mN{_\mu^\nu}

\def\cA{{\cal A}}

\def\od{{\overline d}}
\def\og{{\overline g}}
\def\oR{{\overline R}}
\def\r {{r_*}}

\def\M{{\mathbb M}}
\def\N{{\mathbb N}}

\def\S{{\mathbb S}}

\def\ME{\mbox{$\M_{\rm E}$}}
\def\MJ{\mbox{$\M_{\rm J}$}}

\def\Str{\mbox{$\S_{\rm trans}$}}

\begin{document}
\prepno{gr-qc/0503025}{}

\Title{{\lowercase {\it f}}\,$(R)$ theory of gravity and conformal continuations}

\Aunames {K.A. Bronnikov\auth{1,\dag,\ddag} and M.S. Chernakova\auth{\ddag}}

\Addresses{
\addr{\dag}
    {Centre of Gravitation and Fundamental. Metrology, VNIIMS,
            3-1 M. Ulyanovoy St., Moscow 119313, Russia}
\addr{\ddag}
    {Institute of Gravitation and Cosmology, PFUR,
            6 Miklukho-Maklaya St., Moscow 117198, Russia}
}

\Abstract
{We consider static, spherically symmetric vacuum solutions to the equations
of a theory of gravity with the Lagrangian $f(R)$ where $R$ is the scalar
curvature and $f$ is an arbitrary function. Using a well-known conformal
transformation, the equations of $f(R)$ theory are reduced to the ``Einstein
picture'', i.e., to the equations of general relativity with a source in the
form of a scalar field with a potential. We have obtained necessary and
sufficient conditions for the existence of solutions admitting conformal
continuations. The latter means that a central singularity that exists in
the Einstein picture is mapped, in the Jordan picture (i.e., in the manifold
corresponding to the original formulation of the theory), to a certain
regular sphere \Str, and the solution to the field equations may be smoothly
continued beyond it. The value of the curvature $R$ on \Str\ corresponds to
an extremum of the function $f(R)$. Specific examples are considered.}

\section{Introduction}

    Theories of gravity with the Lagrangian $L=f(R)$, where $f$
    is a certain function of the scalar curvature $R$, are one of the
    well-known and important generalizations of Einstein]s \GR\ (in which
    $L=R)$).  Curvature-nonlinear corrections to the Einstein theory
    are known to emerge due to quantum effects of material fields \cite{qua}.
    Different choices of $f(R)$ have been used for solving cosmological
    problems, in particular, corrections to the Einstein Lagrangian,
    proportional to $R^2$, for describing inflation in the early Universe
    \cite{inflation}, and those of the form $R^{-n}$, where $n > 1$, for
    explaining the present-day accelerated expansion of the Universe
    \cite{accel}. Of no lesser importance are the effects of $f(R)$ theories
    for local configurations, such as, e.g., galaxies and black holes.
    There have been attempts to explain the galactic rotation curves
    by $f(R)$-induced modifications of Newton's law \cite{curve} and
    extensive studies of \bh\ properties in this class of theories \cite{bh}
    (see also references cited in the above-mentioned papers).

    There is a well-known conformal mapping from the manifold \MJ\ with the
    metric $g\mn$, where an $f(R)$ theory is initially formulated
    (it is called the Jordan conformal frame, or Jordan picture), to the
    manifold \ME\ with the metric $\og\mn = g\mn/F(x)$ (the Einstein
    picture), in which the equations of the original theory turn
    into the equations of \GR\ with a scalar field $\phi$ with a certain
    potential $V(\phi)$ (see, e.g., \cite{sok} and references therein).
    If the conformal factor $F(x)$ is everywhere regular, then the basic
    physical properties of the manifolds \MJ\ and \ME\ coincide since,
    in such transformations, a flat asymptotic in \MJ\ maps to a flat
    asymptotic in \ME, a horizon to a horizon, a centre to a centre.
    Using this transformation, some general properties of vacuum \ssph\
    solutions of an arbitrary $f(R)$ theory have been established
    \cite{vac3}. However, of special interest are the cases when a
    singularity in \ME\ maps (due to the properties of $F(x)$) to a regular
    surface in \MJ, and then \MJ\ may be continued in a regular manner
    beyond this surface (this phenomenon has been named a conformal
    continuation \cite{vac3}), and the global properties of the manifold
    \MJ\ can be much richer than those of \ME. The new region may, in
    particular, contain  a horizon or another spatial infinity.

    From a more general viewpoint, the possible existence of conformal
    continuations may mean that the observed Universe is only a region of
    a real, greater Universe which should be described in another, more
    fundamental conformal frame. Detailed discussions of the physical
    meaning and role of different conformal frames may be found in
    Refs. \cite{fara,bm-erice}.

    Necessary and sufficient conditions for the existence of conformal
    continuations (CC) in \ssph\ solutions of scalar-tensor theories of
    gravity have been obtained in Ref.\,\cite{vac4}. In this paper, a
    similar problem is solved for  $f(R)$ theories is space-times of
    arbitrary dimension $D\geq 3$, and two specific examples are considered.

\section{Field equations}

    Consider a theory of gravity with the gravitational field action
\beq
     S_{\rm HOG} = \int d^D x \sqrt{|g|}f(R)               \label{act-hog}
\eeq
    where $f$ is a function of the scalar curvature $R$ calculated for the
    metric $g\mn$ of a space-time $\MJ = \MJ [g]$. In accord with the weak
    field limit $f\sim R$ at small $R$, we assume $f(R) >0$ and $f_R \equiv
    df/dR >0$, at least in a certain range of $R$ including $R=0$, but admit
    $f_R < 0$ and maybe $f < 0$ in general. The vacuum field equations in
    the theory (\ref{act-hog}) are fourth-order in $g\mn$:
\beq
    (R\mN + \nabla_\mu \nabla^\nu - \delta\mN \DAL\,) f_R
                - \half\delta\mN f(R) =0,      \label{eq-hog}
\eeq
    where $\DAL = g\MN \nabla_\mu \nabla_\nu$ and $f_R \equiv df/dR$.

    The conformal mapping $\MJ \mapsto \ME$ with
\beq
    g\mn = F(\phi) \og\mn,\cm  F= |f_R|^{-2/(D-2)},           \label{trans}
\eeq
    transforms the ``Jordan-frame'' action (\ref{act-hog}) into the
    Einstein-frame action
\beq                                                  \label{act-E}
    S = \int d^D x \,\sqrt{|g|} [\oR + (\d\phi)^2 - 2V(\phi)]
\eeq
    where
\bear
    \phi\eql \pm \sqrt{\frac{D-1}{D-2}}\log |f_R|,         \label{phi-hog}
\\
    2V(\phi) \eql |f_R|^{-D/(D-2)} (R|f_R| - f).               \label{V-hog}
\ear
    The field equations due to (\ref{act-hog}) after this substitution
    turn into the field equations due to (\ref{act-E}). Let us write them
    down for \ssph\ configurations, taking the metric $\og\mn$ in the form
\beq                                                          \label{ds_E}
    ds_E^2 = \og\mn dx^\mu dx^\nu =
        A(\rho) dt^2 - \frac{d\rho^2}{A(\rho)} - r^2(\rho) d\Omega_\od{}^2,
\eeq
    where $d\Omega_\od{}^2$ is the linear element on a sphere $\S^{\od}$
    of unit radius, and $\phi=\phi(\rho)$. Three independent combinations of
    the Einstein equations can be written as
\bear
      (A_\rho r^\od)_\rho \eql - (4/\od)r^\od V;                \label{00E}
\\
    \od r_{\rho\rho}/r  \eql -{\phi_\rho}^2;                    \label{01E}
\\
    A (r^2)_{\rho\rho} - r^2 A_{\rho\rho}
                + (\od-2)r_\rho(2Ar_\rho - A_\rho r)\eql 2(\od-1);
                                                \label{02E}
\ear
    where the subscript $\rho$ denotes $d/d\rho$. The scalar field equation
    $(Ar^\od \phi')' = r^\od V_\phi$ follows from the Einstein equations.
    Given a potential $V(\phi)$, (\ref{00E})--(\ref{02E}) is a determined
    set of equations for the unknowns $r,\ A,\ \phi$.

    The metric $g\mn = F\og\mn$ will be taken in a form similar to
    (\ref{ds_E}):
\beq                                                           \label{ds_J}
    ds_J^2 = g\mn dx^\mu dx^\nu =
        \cA(q) dt^2 - \frac{dq^2}{\cA(q)} - \r^2(q) d\Omega_\od{}^2,
\eeq
    The quantities in (\ref{ds_J}) and (\ref{ds_E}) are related by
\beq
            \cA(q) = F A(\rho),                               \label{to_q}
     \cm  \r^2 (q) = Fr^2 (\rho),
     \cm        dq = \pm F d\rho.
\eeq

    Three different combinations of \eqs (\ref{eq-hog}) have the form
\bear
    \od f_R \frac{\r''}{\r} + f_R{}'' \eql 0,                 \label{01J}
\\
    \biggl[ \frac{\od-1}{\r^2} + \frac{\od+2}{2} \r\r'B'
        + \Half \r^2 B''\biggr] f_R                           \label{02J}
            + \Half \r^2 B' f_R{}' \eql 0,
\\
    \bigl[B''\r^2 + (\od+4) B'\r\r'
                + 2(\od+1)B (\r\r'' + \r'^2)\bigr]f_R \inch\cm &&
\nnv
    - [ 2(\od+1) B\r\r' + B'\r^2 ] f_R{}' + f \eql 0,         \label{11J}
\ear
    where the prime stands for $d/dq$ and $B(q) \equiv \cA/\r^2$.
    One can notice that \eqs (\ref{02J}) and (\ref{11J}) [which are
    the difference ${0\choose 0}-{2\choose 2}$ and the ${1\choose 1}$
    component of \eqs (\ref{eq-hog})] are only third-order with respect to
    $g\mn$ while \eq (\ref{01J}) [the difference ${0\choose 0}-{1\choose
    1}$] is fourth-order. \eq (\ref{01J}) is a consequence of (\ref{02}) and
    (\ref{11}).

    In both metrics (\ref{ds_E}) and (\ref{ds_J}) we have chosen the
    ``quasiglobal'' radial coordinates \cite{vac1} ($\rho$ ¨ $q$,
    respectively), which are convenient for describing Killing horizons:
    near a horizon $\rho=\rho_h$, the function $A(\rho)$ behaves as
    $(\rho-\rho_h)^k$ where $k$ is the horizon order: $k=1$ corresponds to
    a simple, Schwarzschild-type horizon, $k=2$ to a double horizon, like
    that in an extremal Reissner-Nordstr\"om \bh\ etc. The function $\cA(q)$
    plays a similar role in the metric (\ref{ds_J}).

\section{Conformal continuations: necessary conditions and properties}

    Let us consider the possible situation when the metric $\og\mn$ is
    singular at some value of $\rho$ while the metric $g\mn$ at the
    corresponding value of $q$ is regular. In such a case $\MJ$ can be
    continued in a regular manner through this surface (to be denoted
    $\Str$), i.e., by definition \cite{vac3,vac4}, we have a conformal
    continuation (CC).

    In our case of spherical symmetry, the sphere $\Str \in \MJ$ may be
    either an ordinary sphere, at which both metric coefficients $\r^2$ and
    $\cA$ are finite (we label such a continuation CC-I), or a Killing
    horizon at which $\r^2$ is finite but $\cA =0$ (to be labelled CC-II).

    Without loss of generality, we suppose for convenience that at $\Str$
    the coordinate values are $\rho = 0$ and $q = 0$ and $\rho > 0$
    in $\ME$ outside $\Str$. According to
    (\ref{to_q}), we must have, in terms of $\og\mn$,
\beq
     F^{-1}\sim r^2 \to 0                               \label{r0}
                \cm {\rm as} \qquad \rho\to 0,
\eeq
    and, in addition, $A(\rho) \sim r^2(\rho)$ for CC-I, whereas for CC-II
    we must have in $\MJ$: $\cA (q) \sim q^n$ at small $q$, where $n\in \N$
    is the order of the horizon.

    Let us use the field equations in $\ME$ for some further estimates.
    \eq (\ref{02E}) may be rewritten in the form
\beq                                                       \label{02E'}
    \frac{d}{d\rho}\biggl(r^D \frac{dB}{d\rho}\biggr) = -2(\od-1)r^{\od-2},
\eeq
    where the function $B(\rho) = A/r^2 = B(q) = \cA/\r^2$ is invariant
    under the transformation (\ref{trans}) and should be finite at
    $\rho = q = 0$. Moreover, since $\Str$ is a regular sphere in $\MJ$,
    $B(q)$ should be a smooth function near $q=0$, and in its Taylor
    series expansion the first two terms may be written as
\beq
     B(q) = B_0 + B_n q^n + \ldots, \cm n \in \N,            \label{B-an}
\eeq
    where $B_0 \ne 0$ for CC-I and $B_0 = 0$ for CC-II.

    Without knowing the potential $V(\phi)$, we cannot specify the behaviour
    of the functions $B(\rho)$ and $r(\rho)$, we only have a relation
    between them given by \eq (\ref{02E'}). There, we should require
    $1/r^2 = o(1/\rho)$, otherwise (\ref{to_q}) would give an infinite value
    of $q$ at $\rho=0$. Taking, for simplicity, $r(\rho) \sim \rho^m$,
    $m>0$, we then have to require $m < 1/2$. Some general restrictions
    can be obtained.

    Let us first consider 3D gravity, $D=3$. Then \eq (\ref{02E'}) gives
    $dB/d\rho = c_1/r^3$, $c_1 = \const$. In case $c_1=0$ we have $B=\const$
    which agrees with (\ref{B-an}). If $c_1\ne 0$, assuming, as before,
    $r\sim \rho^m$ ($0 < m < 1/2)$, we obtain $q \sim \rho^{1-2m}$ and
    $B(\rho) \approx B_0 + B_1 \rho^{1-3m}$ ($B_0, B_1 = \const$) and have
    to put $m < 1/3$ in order to have $B-B_0 \to 0$ as $\rho\to 0$. Lastly,
    comparing this expression for $B$ with (\ref{B-an}), we arrive at $n =
    (1-3m)/(1-2m) < 1$, which contradicts (\ref{B-an}).  We conclude that
    {\it for $D=3$, a CC can only exist if $B = B_0 = \const$}.

    For $D > 3$, let us again assume $r\sim \rho^m$ ($0 < m < 1/2)$, so that
    $q \sim \rho^{1-2m}$. Then, according to (\ref{B-an}), $B - B_0$ behaves
    at small $\rho$ as $\rho^{n(1-2m)}$. On the other hand,
    \eq (\ref{02E'}) now gives
\beq                                                           \label{D>3}
    \frac{d}{d\rho} [\rho^{Dm + n(1-2m) -1}] \sim \rho^{m(\od-2)}.
\eeq
    The exponent inside the square brackets is nonzero, since otherwise we
    would have $n = (1-Dm)/(1-2m) <1$ for $D > 2$. So, differentiating and
    comparing the exponents on the two sides of (\ref{D>3}), we obtain
    (recall that $D= \od+2$) $2-4m = n (1-2m)$ whence $n=2$, in agreement
    with (\ref{B-an}).

    We conclude that, for $D \geq 4$, a CC is possible with
    $r\sim \rho^m$ ($0 < m < 1/2)$, and in this case the function
    $B(q)$ behaves near $\Str$ as
\beq
        B(q) = B_0 + \half B_2 q^2 + o(q^2), \cm B_2 \ne 0.      \label{B_}
\eeq
    Moreover, substituting $B-B_0$ from (\ref{B_}) to \eq (\ref{02E'}), we
    see that its left-hand side behaves as $B_2 \rho^{(D-4) m}$, having the
    sign of $B_2$, whereas its right-hand side has the same $\rho$
    dependence but is negative. Therefore we have to put $B_2 < 0$, which
    means that {\sl the function $B(q)$ has a maximum at $q=0$.}

    All this was obtained by comparing the metrics $g\mn$ and $\og\mn$,
    {\sl without specifying a theory in which the CC takes place, and for
    both kinds of transitions, CC-I and CC-II. Both kinds of transitions are
    thus possible for $D > 3$, and, in particular, in CC-II $\Str$ is a
    double horizon connecting two T regions } (since $B = \cA/\r^2$ is
    negative at both sides of $\Str$).

    In 3D gravity only CC-I are admissible: a horizon, at which $B=0$ but
    $B\ne 0$ in its neighbourhood, is inconsistent with the condition $B =
    \const$.

    Now, for the theory (\ref{act-hog}), the CC conditions can be made more
    precise. A transition surface $\Str$ should correspond to values of $R$
    at which the function $F(\phi)$ introduced in the transformation
    (\ref{trans}) tends to infinity, i.e., where $f_R =0$. In this case,
    according to (\ref{r0}), near $\rho=0$ we have $f_R^{-2/\od} \sim
    r^{-2}$, whence $f_R \sim r^\od$. Substituting this to the expression for
    $\phi$ in (\ref{phi-hog}), we obtain
    $\pm \phi \approx \sqrt{\od(\od+1)} \ln r$. Then, excluding $\phi$ from
    \eq (\ref{01E}) and integrating it under the initial condition $r(0)=0$,
    we finally obtain
\beq                                                           \label{r_)}
    r \approx \const\cdot \rho^{1/D}\cm {\rm as}\quad \rho\to 0.
\eeq
    This result justifies our previous assumption $r \sim \rho^m$ [moreover,
    the value $m = 1/D$ is in the allowed range $(0,1/2)$]. According to
    (\ref{to_q}), we also obtain
\beq                                                           \label{F,q_}
     F \sim \rho^{-2/D},\cm q\sim \rho^{1-2/D} \cm {\rm as}\quad \rho\to 0.
\eeq

    We can now discuss how the curvature $R$ in $\MJ$ is changing across
    $\Str$. We know that if $R = R_0$ at $\Str$, then $f_R (R_0) =0$, but
    we do not know which is the first nonzero derivative of $f(R)$ at
    $R=R_0$. An inspection shows that if $f_{RR}(R_0) \ne 0$, i.e.,
    $f_R \sim R-R_0 \sim \rho^{\od/D} \sim q$ near $\Str$, therefore
    $dR/dq \ne 0$: the curvature changes smoothly and passes from the range
    $R < R_0$ to the range $R > R_0$ or vice versa. On the other hand, if we
    assume that $f_R \sim (R-R_0)^p$, $p > 1$, then $dR/dq \to \infty$ as
    $q\to 0$, i.e., there is no smooth transition. We conclude that {\sl a
    CC is only possible at such values of $R$ where $f(R)$ has an extremum
    with\/} $f_{RR} \ne 0$.

    The above results hold for both CC-I and CC-II, if any. For CC-I we
    know, in addition, that $A(\rho) \sim r^2(\rho) \sim \rho^{2/D}$ at
    small $\rho$.

    Summing up, we have the following necessary conditions and properties of
    a CC at $\rho=q=0$ for a \ssph\ configuration in the theory
    (\ref{act-hog}) in $D\geq 3$ dimensions:
\begin{description}
\item[(a)]
    $f(R)$ has an extremum, at which $f_R = 0$ and $f_{RR}\ne 0$;
\item[(b)]
    $dR/dq \ne 0$ at $q=0$, hence the ranges of the curvature $R$ are
    different at the two sides of $\Str$;
\item[(c)]
    in the Einstein frame, $r(\rho) \sim \rho^{1/D}$ as $\rho\to 0$;
\item[(d)]
    in the Jordan frame, $B(q)$ behaves at small $q$ according to (\ref{B_})
    with $B_2 < 0$, i.e., has a maximum at $q=0$.
\item[(e)]
    For $D=3$, $B(\rho) = B(q) =\const$;
\item[(f)]
    A CC-II is only possible for $D\geq 4$, and $\Str$ is then a double
    horizon connecting two T regions.
\end{description}

\section{Sufficient conditions for conformal continuations}

\subsection{CC-I: continuations through an ordinary sphere}

    Let us prove that the above necessary conditions for CC-I are also
    sufficient. In other words, given a theory (\ref{act-hog}) with a
    smooth function $f(R)$ such that $f_R =0$ and $f_{RR} \ne 0$ at some
    $R=R_0$, there exists a solution to the field equations which is
    smooth in a neighbourhood of the sphere $\Str$ ($R=R_0$).

    It is sufficient to show that there is a solution to the field equations
    (\ref{01J})--(\ref{11J}) in the form of Taylor series near $q=0$.
    It is convenient to take $B(q)$ and $s(q)=\r^2 (q)$ as the two unknown
    metric functions and to find them from \eqs (\ref{02J}) and (\ref{11J}),
    rewritten as
\bear
    \Bigl[ 4(\od-1) + (\od + 2) B'ss' + 2 B'' s^2\Bigr] f_R   \label{02}
                + 2 B' s^2\, f_R{}' \eql 0,
\yy
    \bigl[B''s + \half (\od+4) B's'
                + (\od+1)B s'')\Bigr] f_R 
         - [ (\od+1) B s' + B' s ] f_R{}' + f \eql 0;         \label{11}
\ear
    as before, the prime denotes $d/dq$.

    We seek a solution in the form
\bear
     s(q) \eql \sum_{k=0}^{\infty} \frac{1}{k!} s_k q^k,       \label{s(q)}
\\
     B(q) \eql \sum_{k=0}^{\infty} \frac{1}{k!} B_k q^k.       \label{B(q)}
\ear
    The scalar curvature $R$ is a known function of $B$, $s$ and
    their derivatives:
\beq
     R\,s = \od(\od-1) - B''s^2 - (\od+2) B'ss'
            - (\od+1) B s s'' - \qua \od(\od+1) Bs'^2
                    \vphantom{\frac{1}{4}}    \label{eq-R}
\eeq
    It is helpful, however, to treat $R(q)$ as one more unknown function and
    then to use (\ref{eq-R}) as one more field equation. Accordingly, $R(q)$
    is sought for in the form
\beq
     R(q) = \sum_{k=0}^{\infty} \frac{1}{k!} R_k q^k.         \label{R(q)}
\eeq

    \eqs (\ref{02}) and (\ref{11}) contain the function $f(R)$ from the
    original action (\ref{act-hog}), which can be specified as a Taylor
    series near $R = R_0$,
\beq
     f(R) = \sum_{n=0}^{\infty} \frac{1}{n!} f_n (R-R_0)^n,   \label{f(R)}
\eeq
    with known coefficients $f_n$. We assume that at small $R$ the theory is
    close to GR, i.e., $f(R)$ is approximately a linear function with
    nonzero slope, whereas at a CC we have $f_R=0$. Therefore we assume $R_0
    \ne 0$.

    In accord with the necessary conditions and properties of CCs
    found in the previous section, we require
\bearr
      s_0 > 0, \cm B_0 \ne 0, \cm B_1 =0, \cm B_2 < 0,
\nnn
      R_0 \ne 0; \cm R_1 \ne 0, \cm f_1 =0, \cm f_2 \ne 0.   \label{><0}
\ear

    In the expressions for $f(R)$ and its derivatives, one has to
    substitute a series into a series; it is still possible, however, to
    single out in each order of magnitude $O(q^n)$ a senior term,
    containing the coefficient $R_k$ with the largest $k$. One evidently has
    $(R - R_0)^n = (R_1 q)^n + o(q^n)$, therefore it is easy to see that
    the $R_n$ with greatest $n$ appear in $q$-expansions from the lowest
    powers of $R-R_0$. The function (\ref{f(R)}) with substituted $R(q)$ is
    represented by the following expansion:
\bear                                                         \label{f(q)}
    f(R) \eql f_0 + \Half f_2
     \biggl( R_1 q + \ldots + \frac{R_n}{n!}q^n + \ldots \biggr)^2 + \ldots
\nn
     \eql f_0 + \Half f_2 R_1^2 q^2 + \ldots +
          \biggl(\frac{f_2 R_1}{(n-1)!} R_{n-1} + K_{n-1}\biggr) q^n +\ldots.
\ear
    where $K_{n-1}$ is a certain combination of $R_0, \ldots, R_{n-1}$,
    whose explicit form is insignificant for us.

    In a similar way, we obtain for the derivatives of $f(R)$:
\bearr                                                         \label{f_R}
     f_R = \sum_{k=0}^{\infty} \frac{1}{k!} f_{k+1} (R-R_0)^k
         = f_2 R_1 q + \half (f_2 R_2 + f_3 R_1^2) q^2 + \ldots
            + \biggl(\frac{f_2}{n!}R_n + K_{n-1}\biggr)q^n +\ldots,
\nnn
     f_R{}' = f_2 R_1 + (f_2 R_2 + f_3 R_1^2)q + \ldots
            + \biggl(\frac{f_2}{n!}R_{n+1} + K_n\biggr)q^n
        +\ldots,
\ear
    and so on, with the same meaning of $K_{n-1},\ K_n$.

    Now, let us assume that the following constants (initial
    data) are given:
\beq                                                        \label{give}
     s_0 > 0, \cm B_0 \ne 0, \cm B_1 =0, \cm R_0 \ne 0, \cm R_1 \ne 0.
\eeq

    A further consideration splits into four cases.

\medskip\noi
    {\bf 1. General case:} $D > 3$, $f_0 \ne 0$.
    We obtain in the order of magnitude $O(q^0)$:
\bearr
    (\ref{02})[0]:\cm \mbox{holds automatically};           
\nnnv
    (\ref{11})[0]:\cm f_2 R_1 (\od+1) B_0 s_1 = f_0;
\nnnv
    (\ref{eq-R})[0]:\cm  R_0 s_0 + B_2 s_0^2 + (\od+1)B_0 s_0 s_2
                = \od(\od-1) - \qua \od(\od+1) B_0 s_1^2.
\earn
    From (\ref{11})[0] we express $s_1$ in terms of known constants
    while (\ref{eq-R})[0] connects $B_2$ and $s_2$ with known constants
    including $s_1$ just found.

    In the order $O(q^1)$ we obtain
\bearr                                                      
    (\ref{02})[1]: \cm B_2 s_0^2 = - (\od-1),
\nnnv
    (\ref{11})[1]: \cm (\od+1)(f_3 R_1^2 + f_2 R_2) B_0 s_1=0,
\nnnv
    (\ref{eq-R})[1]: \cm B_3 s_0 + (\od+1) B_0 s_3 = ...,
\earn
    where the dots in (\ref{eq-R})[1] denote a combination of previously
    known quantities including $B_2$ and $s_2$. Note that $B_2 <0$ is
    obtained automatically.

    According to (\ref{11})[0], we have $s_1\ne 0$, hence (\ref{11})[1] leads
    to  $f_3 R_1^2 + f_2 R_2 =0$, so that $R_2$ is expressed in terms of
    $R_1$, $f_2$ and $f_3$. It also means that the quantity $f_R{}''$ (see
    (\ref{f_R}) is zero at $q=0$.

    In further orders $O(q^n)$, $n > 1$ the equations give
\bearr                                                      \label{eq-n}
    (\ref{02})[n]: \cm (n+1)R_1 B_{n+1} =\ldots,
\nnnv
    (\ref{11})[n]: \cm
        (n-1) R_1 s_0 B_{n+1} + (n-1)(\od+1) R_1 B_0 s_{n+1}
               - (d+1) B_0 s_1 R_{n+1} = \ldots,,
\nnnv
    (\ref{eq-R})[n]: \cm
            s_0 B_{n+2} + (\od+1) B_0 s_{n+2} = \ldots,
\ear
    where, as before, the dots denote combinations of known quantities and
    Taylor coefficients from previous terms of the expansions.

    We are now ready to formulate an algorithm for consecutively finding all
    $B_n,\ s_n,\ R_n$. We have seen that using the four equations
    (\ref{11})[0]--(\ref{11})[1] we have found the sought-for
    coefficients up to number 2. Now, assuming that we know these
    coefficients up to number $n$, we obtain $B_{n+1}$ from (\ref{02})[$n$],
    then, using it, we find $s_{n+1}$ from (\ref{eq-R})[$n-1$] and lastly
    $R_{n-1}$ from (\ref{11})[$n$].

    The existence of this algorithm proves the existence of a smooth
    solution to \eqs (\ref{01J})--(\ref{11J}), i.e., of the conformal
    continuation.

\medskip\noi
    {\bf 2.} $D > 3,\ f_0 = 0$. We obtain the same relations as in case 1 in
    the orders $O(1)$ and $O(q)$, with the only difference that now \eq
    (\ref{11})[0] leads to $s_1 = 0$, to be taken into account in further
    equations. \eq (\ref{02})[1] gives $B_2$, and after that (\ref{eq-R})[0]
    expresses $s_2$ in terms of known quantities, while \eq (\ref{11})[1]
    holds trivially.

    In higher orders of magnitude, $n\geq 2$, we obtain the following
    equations:
\bearr                                                      
    (\ref{02})[n]: \cm
    s_0^2 (n+1) R_1 B_{n+1} - (n-1)(\od-1) R_n = \ldots,
\nnnv
    (\ref{11})[n]: \cm
    R_1 s_0 B_{n+1} + (\od +1) R_1 B_0 s_{n+1}
                 - [s_0 B_2 + (\od+1) B_0 s_2] R_n = \ldots,
\nnnv
    (\ref{eq-R})[n-1]: \cm   s_0 B_{n+1} + (\od +1) B_0 s_{n+1} = \ldots.
\earn

    One can see that, as in case 1, there is a recursive algorithm of
    finding all $B_n,\ s_n,\ R_n$, provided the expression
\beq                                                            \label{**}
    s_0 B_2 + (\od+1) B_0 s_2 = -R_0 + \frac{1}{s_0}\od(\od-1)
\eeq
    is nonzero. Indeed, given $R_{n-1},\ B_n,\ s_n$ from previous orders,
    we find a combination of $B_{n+1}$ and $s_{n+1}$ from
    (\ref{eq-R})$[n-1]$, substitute it into (\ref{11})[$n$] to obtain
    $R_n$, then (\ref{02})[$n$] gives us $B_{n+1}$, and, knowing it, we
    lastly obtain $s_{n+1}$ from (\ref{eq-R})[$n-1$].

    If the quantity (\ref{**}) is zero, we generally obtain two different
    expressions for the combination in the left-hand side of
    (\ref{eq-R})[$n-1$], i.e., the set of equations is, in general,
    inconsistent. We can conclude that the sought-for solution exists in all
    cases with this exception.

\medskip\noi
    {\bf 3.} $D=3,\ f_0 \ne 0$. Now, a necessary condition for a CC is that
    $B = B_0 = \const$, which considerably simplifies the equations. Thus,
    \eq (\ref{02}) now becomes trivial, while the other equations lead to
    consecutive determination of all $s_n$ and $R_n$ just as in case 1.

\medskip\noi
    {\bf 4.} $D=3,\ f_0=0$. The situation is like that in case 2 but
    simpler. \eq (\ref{02}) holds trivially; we again obtain $s_1=0$ from
    (\ref{11})[0], then $s_2\ne 0$ from (\ref{eq-R})[0] and $R_2$ from
    (\ref{01J})[0]. Then, (\ref{11})[1] holds automatically, (\ref{eq-R})[1]
    gives us $s_3$, and then, for $n\geq 2$, knowing $R_{n-1}$ and $s_n$, we
    obtain the next coefficients $s_{n+1}$ from \eq (\ref{eq-R})[$n-1$] and
    $R_n$ from (\ref{11})[$n$], containing the combination
    $R_1 s_{n+1} - s_2 R_n$.

    Thus the continuation CC-I exists in all cases described by the necessary
    conditions of the previous section, with the only exception that $D >3$
    and the quantity (\ref{**}) is zero.

\subsection{CC-II: continuations through a horizon}

    For a CC-II, $B_0=0$, and the expansion of $B(q)$ in (\ref{s(q)})
    begins with $k=2$. We now specify the following constants:
    $s_0>0$, $B_0 = B_1 = 0$, $R_0 \neq 0$, $R_1 \neq 0$.  The existence of
    a smooth solution to \eqs (\ref{01J})- (\ref{11}), i.e., the
    existence of a conformal continuation, is proved in a way quite similar
    to CC-I. Recall that a CC-II can only exist for $D > 3$. Besides, CC-II
    are evidently a phenomenon of very special nature because a double
    horizon is a very special case of a sphere in a \sph\ space-time.

    Strictly speaking, the above proofs only provide the existence of
    solutions in the form of asymptotic series, and it is hard to study
    their convergence in a general form. We would note, however, that
    equations with analytical coefficients have, in general, analytical
    solutions, whose Taylor expansions have finite convergence radii near
    their regular points.

\section{Examples}

    Consider two simple examples of exact solutions to \eqs
    (\ref{02}), (\ref{11}), (\ref{eq-R}) with CC-I continuations
    in space-time with the dimensions $D=3$ and $D=4$.

\medskip\noi
    {\bf Three-dimensional example.} In case $D=3$, \eq (\ref{02}) holds
    automatically while (\ref{11}) and (\ref{eq-R}) (taking into account
    that $f'_R = f_{RR}R'$) are written in the form
\bearr                          \label{111}
    2Bs''-2Bs'R' f_{RR}+f=0,\cm R=-2Bs''-Bs'^2/2s.
\ear
    These are two equations for three unknowns $R$, $s$ and $f$, so one of
    them may be taken arbitrarily.  Let $s = q^k$ (choosing the units
    accordingly) with $k\neq 0,\ 2$ (since otherwise we would obtain $R =
    \const$). Then the second equation (\ref{111}) gives $R = -\half Bk (5k
    - 4) q^{k-2}$ while the first one takes the form
\bear
     4(k-2) R^2\,f_{RR} - 4(k-1)R\,f_R + (5k-4) f=0.
\ear
    Its solution
\beq                    \label{333}
     f=C_1R^{x_1}+C_2R^{x_2},\cm
                  x_{1,2}=\frac{2k-3 \pm \sqrt{-k^2+2k+1}}{2(k-2)}
\eeq
    is real for $(1-\sqrt{2}) \leq k \leq (1+\sqrt{2})$. Let us specify
    the numerical values of the constants: let there be $k=12/5$, then
    $x_1 = 5/2$, $x_2 = 2$, $R= - (48/5) B q^{2/5}$. Let us take, further,
    $C_1 = (2/5) (-48 B/5)^{-3/2}$, $2C_2 = (48 B/5)^{-1}$ (recall that
    $B <0$). Then the condition $f_R=0$, corresponding to the transition
    sphere \Str\ in a CC, holds at $q= q_{\rm trans} = 1$.

    The metric in the Jordan and Einstein pictures has the form
\beq
     ds_J^2 = B q^{12/5}dt^2 - B^{-1}q^{-12/5}dq^2-q^{12/5}d\Omega^2,
 \cm
     ds_E^2 = q^{4/5} (q^{1/5}-1)^2 ds_J^2.
\eeq
    The Jordan metric is singular at $q=0$, while the Einstein metric is
    singular at $q=0$ as well as at $q= q_{\rm trans}=1$.  Thus the single
    manifold $\MJ$ corresponds to two manifolds $\ME_1$ and
    $\ME_2$: one is described by the values $q > 1$, the other by $0 < q
    < 1$. Let us present the expressions for the scalar field and its
    potential in the Einstein picture:
\beq
    \phi=\pm \sqrt{2} \ln |q^{3/5}-q^{2/5}|, \cm
    V =  B (2.4 q^{4/5} - 2.88 q)/ (q^{3/5} - q^{2/5})^3.
\eeq
    Both quantities are monotonic in the regions $0 <q <1$ and $q>1$,
    so that the function $V(\phi)$ is determined.

\medskip\noi
{\bf Four-dimensional example.} One of the solutions to \eqs (\ref{02}),
    (\ref{11}), (\ref{eq-R}) at $D=4$ is given by the functions
\beq
     f = -acR + 2c\sqrt{R} = 2c/q-ac/q^2, \qquad
     s = q^2, \qquad B=(3q-2a)/6q^3, \qquad R=1/q^2,
\eeq
    where $a,c - \const > 0$. Let for convenience $a=1$, $c=1$ (choosing the
    appropriate units). Then $f_R = 0$ at $q = q_{\rm trans} = 1$.

    The Jordan and Einstein metrics are
\bear
    ds_J^2 = \left(\frac{1}{2}-\frac{1}{3q}\right)dt^2-
        \left(\frac{1}{2}-\frac{1}{3q}\right)^{-1}dq^2-q^{2}d\Omega^2,
\cm
     ds_E^2 = |q-1| ds_J^2.
\ear
    Thus the Jordan metric has a form close to Schwarzschild's, it is
    singular at the centre $q=0$ and has a horizon at $q = 2/3$. Its
    asymptotic is non-flat due to a solid angle deficit equal to $2\pi$,
    i.e., it has the same nature as the asymptotic of a global monopole (as
    can be easily seen by changing the coordinates from $t$ and $q$ to
    $\bar t = t/\sqrt{2}$ and $\bar q = q\sqrt{2}$). In \ME, the metric is
    singular at $q=0$ and $q=1$ and contains a horizon at $q=2/3$. The
    manifold $\MJ$ again has two Einstein counterparts $\ME_1$ and $\ME_2$:
    separately for $q > 1$ and $q < 1$. The first of them has a non-flat
    asymptotic as $q \to \infty$ and a naked singularity at the centre
    ($q=1$), the other has two singular centres at $q=0$ and $q=1$,
    separated by a horizon at $q=2/3$.

    The scalar field and its potential in \ME\ have the form
\[
    \phi = \pm \sqrt{3/2}\ln|q-1|, \cm V = -\half q^{-1}(q-1)^{-2}.
\]

    The above examples are of methodological nature and demonstrate
    essential distinctions between the descriptions of the theory in the
    Jordan and Einstein pictures when there is a conformal continuation.

\end{document}